\documentclass[pra,preprint,tightenlines,eqsecnum]{revtex4}

 \DeclareFixedFont{\fiverm}{OT1}{cmr}{m}{n}{5pt}
 \input{prepictex}
 \input{pictex}
 \input{postpictex}

\begin{document}

 \title[Spectrum, Orbits, and Scaling]{Spectral Oscillations,
 Periodic Orbits, and Scaling} 

 \author{S. A. Fulling}
 \email{fulling@math.tamu.edu}
 \homepage{http://www.math.tamu.edu/~fulling}
\affiliation{Mathematics Dept., Texas A\&M University,%
  College Station, TX, 77843-3368 USA}

 \date{revision of February, 2002}

  \begin{abstract}
 The eigenvalue density of a quantum-mechanical system
 exhibits oscillations, determined by 
 the closed orbits of the corresponding classical system;
this relationship is simple and strong for waves in billiards or on 
manifolds, but becomes slightly muddy for a Schr\"odinger equation 
with a potential, 
 where the orbits depend on the energy.
In special cases the simplicity has been restored by rescaling
the size of the orbit, and perhaps a coupling constant at the
same time.  We point out that the goal can be achieved for {\em
any\/} system of this class simply by rescaling the overall
coupling constant.  In each of these situations we inspect
critically the relation between the oscillation frequency and the
{\em period\/} of the orbit; in many cases it is observed that a
characteristic {\em length\/} of the orbit is a better indicator.
 When these matters are properly understood, the periodic-orbit 
theory for generic quantum systems recovers the clarity and 
simplicity that it always had for the wave equation in a cavity.
 Finally, we comment on the alleged ``paradox'' that semiclassical 
periodic-orbit theory is more effective in calculating low energy 
levels than high ones.
\end{abstract}

 \maketitle

 \section{Introduction}\label{sec:intro}

 The basic principle relating classical periodic orbits to quantum 
spectra has been succinctly stated by Delos and Du~\cite{DD}: 
\begin{quotation}
The average density of states as a function of energy [is] equal to 
a smooth monotonic function, related to the volume occupied by the 
energy-shell in phase-space, plus a sum of sinusoidal oscillations. 
 The wavelength and amplitude of each oscillation are respectively 
correlated with the period and the stability of a periodic orbit 
of the system.
 For given energy resolution $\Delta E$, only those periodic orbits 
are significant for which the period is less than 
 $2\pi \hbar/\Delta E$.
\end{quotation} \noindent 
An example of a precise theorem along these lines can be found 
in~\cite{CRR}.
 (Other expository works include \cite{FW,Berry,Bolte,Rob}.)

This principle,
 after a prehistory associated with the mathematicians Poisson and 
Selberg, sprang into being 
  in its general form  approximately 30 years ago,
in three  research programs dealing with three distinct contexts:
 \begin{enumerate}
\item Balian and Bloch \cite{BB3} considered a bounded region 
in space.
 The differential operator 
 (corresponding to a quantum Hamiltonian~$H$) 
is the Laplacian with standard 
 (Dirichlet or Neumann) boundary conditions.
 The relevant classical paths are straight lines with specular 
 reflection at the boundary.
 
 \item Gutzwiller \cite{Gutz3,Gutz4} studied a Schr\"odinger 
equation in infinite space, with a potential that is responsible 
for confining particles to compact regions and creating a discrete 
energy spectrum. The periodic orbits involved are those of the full 
classical dynamics governed by that potential function. 
 (Balian and Bloch also studied this situation in a later 
paper~\cite{BB5}.)
 
 \item The paper \cite{BB3} inspired a sequence of works by
 mathematicians  \cite{ColV,Chaz,DG}
 set in a compact Riemannian manifold without boundary.
 Among the very general operators studied in \cite{DG},
 the case most pertinent for us  is the
  Laplace--Beltrami operator for the manifold plus 
a  potential function that is treated as a 
 {\em perturbation}.
That is, the manifold itself is responsible for confining the orbits
and making the spectrum discrete, and the periodic orbits that 
appear in the theorem are the closed geodesics of the Riemannian 
metric, independent of the potential.
 \end{enumerate}\noindent
Although ultimately the most important thing about these three 
bodies of work is  their similarity,
 there are some characteristic differences among them.

First,
 Gutzwiller works with the Schr\"odinger equation,
\begin{equation}
 i\, \frac{\partial\psi}{\partial t} = H \psi \qquad(\hbar=1),
 \label{1.1}\end{equation}
 the others (except Colin de Verdi\`ere \cite{ColV}) with the
wave equation,                         
\begin{equation}
 -\, \frac{\partial^2\psi}{\partial t^2} = H\psi 
 \label{1.2}\end{equation}
 or its first-order pseudodifferential form
 \begin{equation}
 i\,\frac{\partial\psi}{\partial t} = \sqrt H\, \psi.
  \label{1.3}\end{equation}
 (Indeed, the authors of \cite{DG} go so far as to say that
 ``No statement like [our theorem] holds if we replace the first 
order operator [$\sqrt H$] by a higher order operator.'')
 
 Second, Gutzwiller and the many physicists who have followed him
 usually  speak of periodicities in the {\em energy\/} spectrum, 
determined (reciprocally) by the {\em times\/} (periods) of the 
classical orbits.
 But the conclusions of the other works are formulated in terms of
 periodicity in wave frequency, $\omega$
  (essentially the {\em square root\/} 
of energy, $E$),
determined by the {\em lengths\/} of the orbits. 
 A casual reader  might think that the distinction between 
energy periodicity and frequency periodicity arises from the 
respective concentration on the Schr\"odinger equation or the wave 
equation;
 in other words, that the square root relating $E$ to $\omega$  is 
merely the square root relating the quantum Hamiltonian to the 
pseudodifferential operator $\sqrt H$ appearing in the wave 
equation~(\ref{1.3}). 
 That would be wrong.
 Indeed, in hindsight, it is obviously wrong:
 The conclusions of the theory are statements about the 
spectrum of the {\em same\/}
 elliptic second-order differential operator~$H$.
These facts cannot depend upon which time-dependent equation,
(\ref{1.1}) or (\ref{1.3}),
  was used as a technical tool in discovering them.
 
 A third difference is that the orbits studied in \cite{BB3}
  and \cite{ColV,Chaz,DG} 
are energy-independent objects, but the closed orbits in a 
potential vary with the energy. 
The latter fact means that the parametrization of spectral 
oscillations  by orbits
can be meaningful only over rather short energy 
intervals, in general.
 (An exception is systems with homogeneous potentials,
 where the orbits at different energies are related merely by 
dilation in space.  See Sec.~\ref{sec:hom} and 
 \cite[text surrounding 
(1.40)]{Bolte}.) 
 Friedrich and Wintgen \cite{WF,FW}   
pointed out that this problem can be at least partially avoided by 
broadening the point of view to allow variation of a coupling 
constant with energy.
The primary aim of the present paper is to generalize their 
observation,
 showing that it has several variants, one that is applicable 
to any system, and others that may be more appropriate for 
systems with certain scaling symmetries.
 In each case the natural variable of the spectral oscillations is 
a certain power of the energy.
 It is argued that the identification of the frequencies of these 
oscillations with the {\em periods\/} of the orbits is rather 
forced;
 some other quantity, often a {\em length}, is more pertinent.

These matters occupy most of the remaining sections of this paper.
 Sec.~\ref{sec:peri}
  is a digression to establish a (known) relation among 
action, energy, and period when the coupling constant 
 is {\em not\/} scaled; in the context of scaling, this relation no 
longer holds, but neither is it necessary.
 In Sec.~\ref{sec:wave}
  we elucidate the relationship between the wave 
 (\ref{1.2})--(\ref{1.3}) 
and Schr\"odinger (\ref{1.1}) approaches (and their correlates in 
classical phase space);
 this point is  understood by workers in the field but is 
seldom spelled out.
 (To keep  the treatments of the two approaches as parallel as 
possible, we discuss the wave equation in the terminology  of 
relativistic quantum theory;
 this should not be allowed to obscure the applicability of 
 periodic-orbit analysis in other, more classical, contexts, such 
as optics and acoustics, where quantization is just a metaphor.)
 Finally, Sec.~\ref{sec:reso}
  addresses the frequently remarked-upon  counterintuitive fact 
that the ``semiclassical'' periodic-orbit method is more effective 
in reproducing low-lying eigenvalues than eigenvalues in the regime 
of large quantum numbers.

    {\em Bibliographical remarks:\/}  The scaling analysis of the
diamagnetic Kepler problem has a long history, which has been
reviewed by Hasegawa et al.~\cite{HRW} as well as Friedrich and
Wintgen~\cite{FW}.  Until the (more recent) paper of Creagh and
Littlejohn \cite{CL1} this work concentrated on the separate
eigenspaces of axial angular momentum (as is natural from the
quantum-mechanical point of view), which somewhat complicated and
obscured the connection with the classical orbits.  The paper
\cite{FW} was the inspiration for the present work, after which
the author became aware of intervening publications by Friedrich
\cite{Fr1,Fr2} and Main et al.~\cite{MJT} which come close to
making the main points of the present paper.  The main difference
is that here we carry out the analysis in terms of simple
``active'' transformations --- replacements of orbits by
different, geometrically similar orbits --- whereas in the
earlier work the arguments seem to involve ``passive''
redefinitions of physical units.  (Moreover, those coordinate
transformations were noncanonical, forcing the introduction of a
varying ``effective Planck constant'' --- a concept that the
present author finds more confusing than helpful.)  Our reference
list is not exhaustive; many other relevant papers can be traced
from the ones cited here.
 
 \section{Wave and Schr\"odinger dynamics}\label{sec:wave}

 A Hamiltonian system is defined by a function 
 $H(\mathbf x,\mathbf p)$ of 
$\mathbf x\in\Omega$, a $d$-dimensional region or manifold, and 
$\mathbf p\in\mathbf R^d$.
 Its classical equations of motion are
 \begin{equation}
 \frac{d \mathbf x}{dt} = \nabla_{\mathbf p} H, \qquad
  \frac{d \mathbf p}{dt} = - \nabla_{\mathbf x} H. 
\label{2.1}\end{equation}
 Energy is conserved:  Each trajectory of the system (\ref{2.1})
  remains on a phase-space submanifold 
 \begin{equation}
 H(\mathbf x,\mathbf p) = E \label{2.2}\end{equation}
 of dimension $2d-1$.

 Suppose that $H(\mathbf x,\mathbf p)$ is quadratic in~$\mathbf p$ and 
nonnegative, and consider a new Hamiltonian function 
 \begin{equation}
 h(\mathbf x,\mathbf p) \equiv H(\mathbf x,\mathbf p)^{1/2}.
 \label{2.3}\end{equation} 
 Calling the new time parameter~$\tau$, calculate the resulting 
equations of motion:
 \begin{equation}
 \frac{d\mathbf x}{d\tau} = \nabla_{\mathbf p} h = 
 {\textstyle\frac12}H^{-1/2} \frac{d\mathbf x}{dt}\,, \qquad
\frac{d\mathbf p}{d\tau} = 
 {\textstyle\frac12}H^{-1/2} \frac{d\mathbf p}{dt}\,. 
 \label{2.4}\end{equation}
It follows that the trajectories of the new system in phase space 
are exactly the same as those of the old one, but parametrized 
differently.
 At a fixed energy the two time scales are simply related by
 \begin{equation}
 \frac{d\tau}{dt} = 2\sqrt E. 
 \label{2.5}\end{equation}
 The relation between $\mathbf p$ and the velocity  is changed by 
the same factor.

Consider the very special case $H= \mathbf p^2/2m$.
 Then $\mathbf p$ is a constant vector with magnitude 
 $p=\sqrt{2mE}$,
 and the velocity in system (\ref{2.1}) satisfies
 \begin{equation}
 \mathbf v \equiv \frac{d\mathbf x}{dt} =
  {\mathbf p\over m}\,, \qquad v^2 = {2E\over m}\,, 
 \label{2.6}\end{equation}
while the velocity in system (\ref{2.4}) satisfies
 \begin{equation}
 \mathbf u  \equiv \frac{d\mathbf x}{d\tau} = 
 \frac1{2\sqrt E}\,{\mathbf p\over m}\,, 
 \qquad u^2 = \frac1{2m} = \text{constant}. 
 \label{2.7}\end{equation}
 The trajectories of $H$~dynamics in space-time are, of course,
 those of a nonrelativistic particle: straight lines 
of arbitrary slope depending on energy.
 The trajectories of $h$~dynamics in space-time are 
 {\em independent of energy}, as appropriate for a
 {\em relativistic massless particle}.
 (Clearly, in this context $m$ does not have the physical 
significance of a mass;
 rather, it parametrizes the wave speed as $c=(2m)^{-1/2}$.)
 Passing to a quantum system by the prescription
 \begin{equation}
  \mathbf p \mapsto -i\nabla_{\mathbf x}\,,\qquad 
 E \mapsto i \,\frac{\partial}{\partial\tau}
\label{2.8}\end{equation}
 converts the relativistic energy relation
  $h(\mathbf x,\mathbf p)=E$ 
 to a special case of (\ref{1.3}),
  which is equivalent to the wave equation
\begin{equation}
 \frac{\partial^2\psi}{\partial t^2} =\frac1{2m} \nabla^2\psi 
 \label{2.9}\end{equation}
 supplemented by a condition of positive frequency.
 The nonrelativistic energy relation (\ref{2.2})
  of course becomes the 
free Schr\"odinger equation, a case of (\ref{1.1}).

The energy independence of relativistic trajectories
  leads to a 
very direct relationship between spectral oscillations and the 
 {\em singularities\/} 
 of the fundamental solution of the wave equation 
(\ref{1.3}).
 This is the meaning of the remark of Duistermaat and Guillemin 
quoted in Sec.~\ref{sec:intro}{}.
 An up-to-date exposition of that methodology is given 
in~\cite{SV}.

 Now turn to a Hamiltonian including a potential energy function,
 \begin{equation}
 H(\mathbf x,\mathbf p) =
  {\mathbf p^2\over 2m} + V(\mathbf x),
  \label{2.10}\end{equation}
 where (for the moment) $V(\mathbf x)$ is nonnegative.  
 The momentum is no longer constant, and the velocities satisfy
 \begin{equation}
 v^2 = \frac2m [E-V(\mathbf x)] 
 \label{2.11}\end{equation}
 (the usual nonrelativistic kinetic-energy relation) and
 \begin{eqnarray}
 u^2 &=&  {p^2 \over 2m(p^2 + 2mV(\mathbf x))}
 \nonumber\\
&=& \frac1{2m}\left( 1-\frac{V(\mathbf x)}E\right)  .
 \label{2.12} \end{eqnarray}
 The trajectories in space-time are now curves, whose slopes at 
each point are greater than they would have been for a free 
particle.
 In particular, in the relativistic case the trajectories stay 
always inside the  local light cones,
  as appropriate for a relativistic 
particle with mass (induced here by an $\mathbf x$-dependent 
interaction). 
 The main point is that the traces of these orbits on configuration 
space are the same as in the nonrelativistic case.

 In passing, note that if $V$ were allowed to be negative, the
 classical relativistic trajectories would be 
``tachyonic'' according to~(\ref{2.12}).
 Nevertheless, the wave equation~(\ref{1.2}) still obeys hyperbolic 
causality.
This situation was studied by Schroer and others 
 in the context of quantum field theory~\cite{Sch}.
 We shall not pursue this case further, because it interferes 
with the definition of the square roots in 
 (\ref{2.3}) and (\ref{1.3}).

 \section{Scaling a billiard}\label{sec:bill}

 Henceforth we adopt the nonrelativistic, or 
 Schr\"odinger-equation, point of view and  restore the 
conventional constant $\hbar$.
  We consider Hamiltonians of the form 
(\ref{2.10})
(without restriction on the sign of~$V$).
  The quantization prescription (\ref{2.8})
  leads unambiguously to a 
second-order elliptic partial differential operator 
 \begin{equation}
 H = -\,\frac{\hbar^2}{2m}\nabla^2 + V(\mathbf x).  
 \label{3.1}\end{equation}
 (In more general cases some variant of Weyl quantization must be 
chosen to resolve operator-ordering ambiguities~\cite{Full}.)
The complete specification of the problems 
 (\ref{1.1})--(\ref{1.3}) may also 
require boundary conditions, which we take to be of the standard
 Dirichlet, Neumann, or Robin type.

     It is assumed that the spectrum of the operator~$H$ is 
     discrete, at least below some threshold.
 The essence of the Gutzwiller trace theory 
\cite{DD,CRR,FW,Berry,Bolte,Rob,BB3,Gutz3,Gutz4,BB5} 
 is that the density of states, $\rho(E)$, contains a term
 \begin{equation}\rho_\gamma(E) = 
 a_\gamma \sin\left[{S_\gamma\over \hbar} + \eta_\gamma\right]
 \label{3.2}\end{equation}
 for each classical periodic orbit $\gamma$ of energy~$E$;
 here $S_\gamma(E)$ is the {\em action\/}
 $\oint \mathbf p\cdot d\mathbf x$ of the orbit~$\gamma$,
 and the details of the amplitude $a_\gamma(E)$ and the phase
 $\eta_\gamma$ do not concern us now.
 (Also, this simplified formulation does not do justice to the 
complications that arise when the orbits are not isolated
and unstable.
We touch on such matters in Sec.\ VIII.)

 Let us examine what (\ref{3.2}) asserts
  for a system of the type studied 
by Balian and Bloch~\cite{BB3}.
 The classical paths are straight lines in a region
  $\Omega\subset \mathbf R^d$
 with specular reflections at the boundary of~$\Omega$.
 (The distinction among different boundary conditions shows up in 
the phases~$\eta_\gamma\,$.)
 In particular, these paths are the {\em same\/} for all~$E$.
 The speed is fixed at $v = \sqrt{2E/m} $
(see (\ref{2.6})).
 If the length of  such a  closed polygonal path, $\gamma$, is 
$L_\gamma\,$,
 then its period is
 \begin{equation}
 T_\gamma(E) = {L_\gamma \over v} = \sqrt{m\over 2E} \, L_\gamma\,. 
\label{3.3}\end{equation}
 The momentum is parallel to the path with fixed magnitude
 $p=\sqrt{2mE}$, so the action is
 \begin{equation}
 S_\gamma(E) \equiv \oint_\gamma \mathbf p\cdot d\mathbf x = 
 \sqrt{2mE} \,L_\gamma = 2ET_\gamma\,. \label{3.4}\end{equation}
 Thus (\ref{3.2}) becomes
 \begin{eqnarray}
 \rho_\gamma(E) &=& a_\gamma \sin\left[{2T_\gamma E\over \hbar} + 
 \eta_\gamma\right]
\label{3.5} \\
 &=& a_\gamma \sin\left[{\sqrt{2m E}\over \hbar}\,L_\gamma
  + \eta_\gamma\right]. 
 \label{3.6}\end{eqnarray}

  It is crucial to note that
$T_\gamma$ is itself a function of~$E$, 
but $L_\gamma$ is constant.
Thus $\rho_\gamma$ in (\ref{3.6}) is 
 {\em exactly and globally\/} 
 (to the extent that $a_\gamma$ and $\eta_\gamma$ can be regarded as 
constant)
 a sinusoidal function of $\omega \equiv \sqrt E$ 
with period
 \begin{equation}
 P_\omega =\sqrt{2\over m} {\pi\hbar \over L_\gamma }\,,
   \label{3.7}\end{equation}
 or $P_\omega={2\pi/L_\gamma}$ if ${\hbar^2\over 2m} =1$, 
 the most natural 
normalization when dealing with the wave equation.
 This is the result obtained by Balian and Bloch~\cite{BB3}
  working from 
the relativistic, or wave-equation, point of view:
 The eigenvalue density oscillates with frequency $L_\gamma$ as a 
function of~$\omega$.

 On the other hand, we have
 \begin{equation}
 \frac{d S_\gamma}{dE} = \sqrt{m\over 2E}\, L_\gamma = T_\gamma
  = {S_\gamma\over 2E}\,. 
 \label{3.8}\end{equation}
 Approximating $S_\gamma(E)$ locally by a linear 
function~\cite{DD},
 one gets from (\ref{3.2}) and (\ref{3.8})
 \begin{equation}
 \rho_\gamma(E) \approx a_\gamma 
 \sin\left[{T_\gamma\over\hbar}(E-E_0)
 +{S_\gamma(E_0)\over\hbar}   + \eta_\gamma\right]. 
 \label{3.9}\end{equation}
 Thus $\rho_\gamma$ oscillates {\em locally\/} as a function of $E$
 with {\em approximately defined\/} frequency 
 $T_\gamma(E_0)/\hbar$
 (not twice that, as might appear from (\ref{3.5})),
 or period
 \begin{equation}
 P_E = {2\pi\hbar\over T_\gamma}\,. 
 \label{3.10}\end{equation}
 This is the sort of statement about spectral oscillations that is 
most common  in the physics literature surrounding the Gutzwiller 
trace formula.

 For completeness we record the time-dependent action
 \begin{equation}
 R_\gamma(T) = S_\gamma(E) - ET_\gamma = ET_\gamma = 
 {mL_\gamma{}\!^2 \over 2T} \,, 
 \label{3.11}\end{equation}
 where
 \begin{equation}
 E = {\textstyle \frac12}mv^2 = {mL^2\over T^2}
  \label{3.12}\end{equation}
 has been used.
 It satisfies
 \begin{equation} 
 \frac{d R_\gamma}{dT} = -E, 
 \label{3.13}\end{equation}
 which is dual to (\ref{3.8}) when (\ref{3.11})
  is interpreted as a Legendre transformation.
 It can also be obtained as
 \begin{equation}
 R_\gamma \equiv \oint_\gamma \mathcal L\, dt = ET, 
 \label{3.14}\end{equation}
 where $\mathcal L$ is the Lagrangian, 
 here equal to the kinetic energy.

\section{Interlude:  The action--period relation}
 \label{sec:peri}

 In converting (\ref{3.2}) to (\ref{3.9}), only the relation
 \begin{equation}
 \frac{d S_\gamma}{dE} = T_\gamma 
 \label{4.1}\end{equation}
 was used.
 This identity is frequently cited as 
 ``a well known theorem of classical mechanics''\negthinspace,
 but a proof is hard to find in the literature.
 Indeed, at first it is not even clear what (\ref{4.1})
  {\em means\/} for a generic Hamiltonian (\ref{2.10}), 
 since $S_\gamma$ was defined for a single closed orbit 
 $\gamma$ existing 
at a particular energy, say $E_0\,$. 
 The discussion in Sec.~\ref{sec:bill} shows that 
 for a billiard the orbit 
indeed persists unchanged (in configuration space) as $E$ varies;
 but when the dynamics involves a potential function the classical 
path defined by an initial point and direction on $\gamma$ will 
generally cease to be a periodic orbit when $E$ deviates 
from~$E_0\,$.
 Generically $\gamma$ will smoothly 
evolve with $E$ into a family of nearby periodic orbits $\gamma(E)$
 (in general, disjoint from $\gamma(E_0)$),
 but even this picture  breaks down at certain singular points
 (e.g., a maximum of the 
potential in dimension~1,
 where two families of orbits merge into one).
 The main point of the remainder of this paper is that such
issues can be evaded by adopting a broader point of view.
 Here, however, we provide a simple derivation of (\ref{4.1}) in 
circumstances where it makes sense.

 {\em Assume\/} the existence of a family $\gamma(E)$ of classical 
closed orbits of (\ref{2.1}) depending smoothly on the energy,~$E$
 (as $E$ varies over some interval, possibly small).
 Assume also that either (1) the period $T_\gamma$ depends smoothly 
and monotonically on~$E$, so that $\gamma$ can be alternatively 
parametrized by~$T$, or (2) $T_\gamma$ is independent of~$E$.
In Case (1) we shall prove
 \begin{equation}
 \frac{d R_\gamma}{dT} = - E, 
 \label{4.2}\end{equation}
 which is equivalent to (\ref{4.1}) under the Legendre transformation
 \begin{equation}
 S = R+ET \qquad \bigl(S= S(E), \quad R= R(T)\bigr). 
 \label{4.3}\end{equation}
 For a family of classical trajectories from a fixed initial 
 space-time point $(\mathbf x',0)$ to a variable final point
$(\mathbf x,T)$
 (and satisfying condition (1)),
 (\ref{4.2}) is the usual 
 (e.g., \cite{Park})
 Hamilton--Jacobi equation for 
 $R$ as a function 
of $(\mathbf x,T)$ (with $E = H(\mathbf x,\nabla R)$);
 but for periodic orbits a different argument is necessary.

 By definition,
 \begin{equation}
 R = \oint_\gamma \mathbf p\cdot d\mathbf x - ET = 
 \int_0^T \mathcal L \, dt, 
\label{4.4}\end{equation}
 \begin{equation}
 \mathcal L = {\textstyle\frac12} m\dot{\mathbf x}(t)^2 
 - V(\mathbf x(t)). 
 \label{4.5}\end{equation}
In the standard way, consider a variation
 \begin{equation}
 \delta R = \int_0^T {\delta \mathcal L\over \delta\mathbf x}\,
  \delta\mathbf x \, dt
 + \bigl.\mathcal L\bigr|_T\, \delta T, 
 \label{4.6}\end{equation}
 where
 \begin{equation}
 \delta\mathcal L = 
 m \dot\mathbf x\cdot \delta\dot\mathbf x 
 -\nabla V\cdot \delta\mathbf x
 = \frac{d}{dt}( m\dot\mathbf x\cdot \delta\mathbf x ) - 
 (m\ddot\mathbf x + \nabla V)\cdot \delta \mathbf x . 
\label{4.7}\end{equation}
 The last term in (\ref{4.7}) vanishes by the equation of motion,
 and integrating the other term yields
 \begin{equation}
 m\dot\mathbf x(T)\cdot\delta\mathbf x(T)
  - m\dot\mathbf x(0)\cdot\delta\mathbf x(0)
 = -m \dot\mathbf x(T)^2 \, \delta T, 
 \label{4.8}\end{equation}
 because
 \begin{equation}
 \mathbf x(T)= \mathbf x(0), \quad 
 \dot\mathbf x(T)= \dot\mathbf x(0), \quad
 (\mathbf x+\delta\mathbf x)(T+\delta T) = 
 (\mathbf x+\delta\mathbf x)(0),
 \label{4.9}\end{equation}
 so that
 \begin{eqnarray}
  \delta\mathbf x(T) &=& (\mathbf x+\delta\mathbf x)(T) - 
 \mathbf x(T) \nonumber\\
 &=& (\mathbf x+\delta\mathbf x)(T+\delta T)
  - (\dot\mathbf x+\delta\dot\mathbf x)(T)\delta T - \mathbf x(0)
 \nonumber\\
 &=& \delta\mathbf x(0) - \dot\mathbf x(T)\,\delta T 
 + O\bigl((\delta T)^2\bigr) .
 \label{4.10}\end{eqnarray}
 Putting all the ingredients into (\ref{4.6}) yields
 \begin{eqnarray*}
  {\delta R\over \delta T} &=&
  -m\dot\mathbf x(T)^2 + {\textstyle\frac12}
 m\dot\mathbf x(T)^2 - V(\mathbf x(T)) \\
 &=& -  {\textstyle\frac12} m\dot\mathbf x(T)^2 - V(\mathbf x(T)) 
\\
 &=& -E, \end{eqnarray*}
 as desired.

 In Case (1),
 (\ref{4.2}) and (\ref{4.3}) imply (\ref{4.1}) by the standard 
Legendre argument.
 In Case (2), the Hamilton--Jacobi equation (\ref{4.2}) does not 
exist, but in that case the foregoing calculation shows that
 $\delta R=0$, and hence (\ref{4.1}) follows from (\ref{4.3}) 
 even more directly.
 In either case, (\ref{3.2})  then   leads as 
before~\cite{DD}  to (\ref{3.9}) 
and (\ref{3.10}), under the additional assumption that $T_\gamma$ is 
sufficiently slowly varying for the concept of a local frequency of 
oscillation of $\rho_\gamma$ to make sense.

 An interesting example 
 (and the principal instance of Case (2))
 is the one-dimensional harmonic oscillator,
 \begin{equation}
 V(x) = {\textstyle\frac12} m\varpi^2 x^2.
  \label{4.11}\end{equation}
 The general solution of its equation of motion is
 \begin{equation}
 x(t) = x_0  \cos {\varpi t} + \frac{p_0}m \,\sin{\varpi t}.
 \label{4.12}\end{equation}
 One calculates
 \begin{equation}
 T = {2\pi\over \varpi}\,,
 \qquad
 E = {p_0{}\!^2\over 2m} + {\textstyle\frac12} m\varpi^2 x_0{}\!^2, 
\label{4.13}\end{equation}
 \begin{equation}
 S = ET = {2\pi E\over \varpi}\,, 
 \qquad
 R = 0. 
 \label{4.14}\end{equation}
 Thus
 \begin{equation}
 \frac{dS}{dE} = T = \frac SE 
 \label{4.15}\end{equation}
 in agreement with (\ref{4.1}) (but in contrast to (\ref{3.8})).

\section{Scaling a coupling constant} \label{sec:coup}

Consider a Hamiltonian
 \begin{equation}
 H(\mathbf x,\mathbf p) = 
 {\mathbf p^2 \over 2m} + \lambda V(\mathbf x) 
 \label{5.1}\end{equation}
 with a generic potential $V$ and a coupling constant~$\lambda$.
 The equations of motion are
 \begin{equation}
 \frac{d \mathbf x} {dt} = \frac {\mathbf p}m \,,\qquad 
 \frac{d \mathbf p} {dt} = - \lambda\nabla V(\mathbf x), 
\label{5.2}\end{equation}
 and any solution curve lies on an energy surface
 \begin{equation}
  {\mathbf p^2 \over 2m} + \lambda V(\mathbf x) = E. 
 \label{5.3}\end{equation}

 Let $\bigl( \mathbf x_0(t), \mathbf p_0(t) \bigr)$ be a solution of
 (\ref{5.2}) and (\ref{5.3}) with $\lambda=\lambda_0$ and $E=E_0\,$.
 For any positive real number~$\alpha$,
 consider
 \begin{equation}
 \mathbf x(t) \equiv \mathbf x_0(\alpha t), \qquad 
 \mathbf p(t) \equiv \alpha \mathbf p_0(\alpha t). 
\label{5.4}\end{equation}
 A short calculation shows that
$\bigl( \mathbf x(t), \mathbf p(t) \bigr)$ satisfies 
  (\ref{5.2}) and (\ref{5.3}) with 
 \begin{equation}
 \lambda=\alpha^2 \lambda_0\,, \qquad E=\alpha^2 E_0\,.  
 \label{5.5}\end{equation}
 In particular, the path $\mathbf x(t)$ in configuration space is 
independent of the parameter $\alpha^2 = E/E_0\,$.
 In this sense the same closed orbit exists for all values of the 
energy.
 When $V=0$ this is the familiar billiard orbit reviewed in 
 Sec.~\ref{sec:bill}.
 When the potential is not zero, however, one must pay the price of 
varying $\lambda$ --- i.e., considering {\em different\/} physical 
systems at different energies ---
 to gain the convenience of fixed orbits.
 (However, in electromagnetic wave problems with dielectrics 
the scaling (\ref{5.5}) {\em does\/} relate solutions of the same 
system~\cite{BKS}.)

 For the orbit (\ref{5.4}) it is clear that
 \begin{equation}
 T = \frac {T_0}{\alpha} \,, 
 \label{5.6}\end{equation}
 and one calculates
 \begin{eqnarray}
 S = \oint_\gamma \mathbf p \cdot d\mathbf x &=&
 \alpha \int_0^T \mathbf p_0 (\alpha t) \cdot 
 [\alpha \dot\mathbf x_0(\alpha t)] \,dt \nonumber\\
 &=& \alpha\int_0^{T_0} \mathbf p_0(\tau) \cdot
  \dot\mathbf x_0(\tau) \, d\tau = 
 \alpha S_0\,.  \label{5.7} \end{eqnarray}
 Thus (\ref{3.2}) becomes (by use of (\ref{5.5}))
 \begin{eqnarray}
 \rho_\gamma(E) &=& a_\gamma \sin\left[ {S_0\over \hbar} \sqrt{E\over 
E_0} + \eta_\gamma\right] \nonumber\\
 &=& a_\gamma \sin\left[ {S_0\over \hbar\omega_0}\,\omega
 + \eta_\gamma\right]. 
 \label{5.8}\end{eqnarray}
 That is, just as for a billiard, we have a globally sinusoidal 
function of $\omega\equiv\sqrt{|E|}$
  with frequency $S_0/\hbar \omega_0\,$, or period
 \begin{equation}
 P_\omega = {2\pi \hbar \omega_0 \over S_0}\,. 
 \label{5.9}\end{equation}
 Thus the analogue of the $L_\gamma$ of a billiard is
(cf.~(\ref{3.7})) 
\begin{equation}
\sqrt{\hbar^2 \over 2m} \, {2\pi \over P_\omega} =   
 {S_0 \over \sqrt{2m |E_0|}} \,. 
 \label{5.10}\end{equation}
 For the harmonic oscillator orbit (\ref{4.12}),
  this characteristic length is 
 $\pi$ times the maximum value of $x(t)$
 (not 4 times, which would be the actual length of the orbit).
 It should be noted that (\ref{5.9}) or (\ref{5.10})
  depends only on the 
orbit~$\gamma$, not on the reference scale arbitrarily chosen to 
correspond to $\alpha=1$.

 One can also look at the local frequency with respect to~$E$,
 following (\ref{3.8})--(\ref{3.10}).
 From (\ref{5.7}) and (\ref{5.5}) one has
 \begin{equation}
 \frac{d S}{dE} = {S \over 2E}\,. 
 \label{5.11}\end{equation}
 But this quantity is no longer equal to~$T$,
 because of the variation of the coupling constant with~$E$.
 For the harmonic oscillator it equals $\frac12 T$, by (\ref{4.15}).
 More generally, since the momentum is parallel to the velocity for 
a system of type (\ref{5.1}), one can compute
 \begin{equation}
 S = \oint_\gamma \mathbf p\cdot \dot\mathbf x \, dt
 = 2 \int_0^T [E- \lambda V(\mathbf x(t))]\, dt,
  \label{5.12}\end{equation}
 and hence
 \begin{equation}
  {S\over 2E} = T - \frac{\lambda}E \int_0^T V(\mathbf x(t))\, dt. 
\label{5.13}\end{equation} 
 Note also that $S/2E$ scales as $1/\alpha$ and therefore cannot be 
regarded as a characteristic time of the orbit~$\gamma$
 (in contrast to the length (\ref{5.10})).

 Finally, because nearby nonperiodic orbits also obey the scaling 
law, the amplitude $a_\gamma$ in (\ref{5.8})
  (determined by the monodromy of 
the Poincar\'e map) is independent of~$E$, whereas the same cannot 
be said for the $a_\gamma$ in (\ref{3.9}).
 The phase shift $\eta_\gamma$ also is constant for Dirichlet and 
Neumann boundary conditions (not Robin).

 Main et al.~\cite{MJT} rescale the kinetic term of the 
Hamiltonian 
instead of the potential term and the total energy.
 Their procedure is equivalent to that of this section together 
with a  redefinition of the  time scale.
 In that approach the energy of a given orbit is always fixed, and 
the quantum parameter (see Sec.~\ref{sec:imp}) is $\sqrt{m}$
 instead of~$\omega$.

 \section{Scaling a homogeneous potential} \label{sec:hom}

 When the potential in (\ref{5.1}) is homogeneous of degree~$\nu$,
 \begin{equation}
 V(\beta \mathbf x) = \beta^\nu V(\mathbf x) 
 \quad(\text{for $\beta>0$}),
  \label{6.1}\end{equation}
 an alternative scaling procedure exists that avoids changing the 
coupling constant.
 Assume that the configuration space~$\Omega$, 
 if not all of~$\mathbf R^d$, 
is invariant under dilations
 (e.g., a half-space or a cone)
 with boundary conditions that are either Dirichlet or pure 
Neumann (not Robin).
(If $\Omega$ does not satisfy this condition,  then it must be dilated 
along with the orbits, and again the system has been replaced by a 
one-parameter family.)
 In place of (\ref{5.4}) consider
 \begin{equation}
 \mathbf x(t) \equiv\alpha^2 \mathbf x_0(\alpha^{\nu-2} t), \qquad 
 \mathbf p(t) \equiv \alpha^\nu \mathbf p_0(\alpha^{\nu-2} t). 
\label{6.2}\end{equation}
 This is a solution of (\ref{5.2}) and (\ref{5.3}) with
  \begin{equation}
 \lambda= \lambda_0\,, \qquad E=\alpha^{2\nu} E_0\,. 
  \label{6.3}\end{equation}
 The orbits for different values of $\alpha$ are not the same, 
 but they are geometrically similar;
 one can say that they are the same if the unit of length is rescaled 
by~$\alpha^2$.

 We now have (parallel to (\ref{5.6})--(\ref{5.8}))
 \begin{equation}
 \alpha = \left( \frac E{E_0} \right) ^{\frac1{2\nu}}, 
 \label{6.4}\end{equation}
 \begin{equation}
 T = \alpha^{2-\nu} T_0\,, \qquad S = \alpha^{\nu+2} S_0\,, 
 \label{6.5}\end{equation}
 \begin{equation}
 \rho_\gamma(E) = a_\gamma \sin\left[{S_0\over \hbar} 
 \left( \frac E{E_0}\right)  ^{\nu+2\over 2\nu} + \eta_\gamma \right]. 
 \label{6.6}\end{equation}
 Thus the variable with respect to which the spectral oscillations 
take place is a peculiar power of~$E$, and the analogue of the 
 characteristic length (\ref{5.10}) is a rather inconvenient
 function of the fiducial action and energy.
 Note that the exponent of $|E|$ in (\ref{6.6}) equals~$1$ for the 
 harmonic oscillator;
 that it acquires the value $\frac12$ (familiar for billiards)
 only as $\nu\to +\infty$;
 that it is negative for $-2<\nu<0$; and
  that the cases $\nu=0$ and $\nu=-2$ are singular.

  The analogue of (\ref{5.11}) is
 \begin{equation}
 \frac{d S}{dE} = {\nu+2\over 2\nu} \, \frac SE\,,
   \label{6.7}\end{equation}
 (\ref{5.12}) remains true, and the analogue of (\ref{5.13}) is
 \begin{equation}
 {\nu+2\over 2\nu}\, \frac SE =
 {\nu+2\over\nu} \left[ T - \frac{\lambda}E 
 \int_0^T V(\mathbf x(t))\,dt\right]. 
\label{6.8}\end{equation}
 On the other hand, the theorem of Sec.~\ref{sec:peri} applies to the orbit 
family (\ref{6.2}), and so
 \begin{equation}
 \frac{d S}{dE} = T.
  \label{6.9}\end{equation}
 Equating (\ref{6.8}) and (\ref{6.9}) yields the identity 
 \begin{equation}
 \int_0^T \lambda V(\mathbf x(t))\, dt = {2ET \over \nu + 2}
  \label{6.10}\end{equation}
 {\em for a periodic orbit of a homogeneous potential of 
degree~$\nu$}, which is a rewriting of the virial 
theorem~\cite{Gold}.

 The Coulomb potential, 
 \begin{equation}
 V(\mathbf x) = - \,{e^2\over r} \qquad 
   (r^2 \equiv x^2 + y^2 + z^2),
 \label{6.11}\end{equation}
 has 
 \begin{equation}
 \nu = -1, \qquad
{\nu+2 \over 2\nu} = -\,\frac12\,.
  \label{6.12}\end{equation}
Furthermore,  in the regime of discrete spectrum and periodic 
orbits of the atom, $E$~is negative.
 Thus (\ref{6.7}) becomes
 \begin{equation}
 \frac{d S}{dE} = + \, {S\over 2|E|}\,, 
  \label{6.13}\end{equation}
 which displays a superficial coincidence with the corresponding 
relation for a billiard, (\ref{3.8}).

\section{Scaling of mixed type} \label{sec:mix}

 A mixture of the strategies of Secs.\ \ref{sec:coup} 
 and~\ref{sec:hom} is appropriate in 
some situations.
 Suppose that the potential is a sum of two terms, each of which is 
homogeneous:
 \begin{equation}
 V_1(\beta \mathbf x) = \beta^{\nu_1} V_1(\mathbf x), \qquad
V_2(\beta \mathbf x) = \beta^{\nu_2} V_2(\mathbf x).
  \label{7.1}\end{equation}
 The equations of motion and the energy equation are
 \begin{equation}
 \frac{d \mathbf x} {dt} = \frac {\mathbf p}m\,,\qquad
 \frac{d \mathbf p}{dt} = - \lambda_1 \nabla V_1(\mathbf x)
  - \lambda_2 \nabla V_2(\mathbf x), 
\label{7.2}\end{equation}
 \begin{equation}
 {\mathbf p^2 \over 2m} + \lambda_1 V_1(\mathbf x) 
 + \lambda_2 V_2(\mathbf x) = E.
 \label{7.3}\end{equation}
 (This is {\em not\/} a perturbative situation:
 Neither potential is assumed to be small compared to the other.)
Following the usual pattern,
  suppose that we have a solution with coupling constants 
  $\lambda_{10}$ and $\lambda_{20}$  and energy $E_0\,$.
 Perform the scaling (\ref{6.2}) appropriate to $V_1\,$:
  \begin{equation}
 \mathbf x(t) \equiv\alpha^2 \mathbf x_0(\alpha^{\nu_1-2} t), \qquad 
 \mathbf p(t) \equiv \alpha^{\nu_1} \mathbf p_0(\alpha^{\nu_1-2} t). 
\label{7.4}\end{equation}
 Then (\ref{7.4}) solves (\ref{7.2}) and (\ref{7.3}) with
 \begin{equation}
 \lambda_1 = \lambda_{10}\,, \qquad 
 \lambda_2 = \alpha^{2(\nu_1-\nu_2)} \lambda_{20}\,,
 \qquad E = \alpha^{2\nu_1} E_0\,.
   \label{7.5}\end{equation}
 Equations (\ref{6.4})--(\ref{6.6}) continue to apply,
  with $\nu=\nu_1\,$.
 
 The point of this transformation is that a fixed orbit can be 
associated with a one-parameter family of classical situations, in 
which one coupling constant is held fixed but the other varies with 
energy.
 The prototype is the Hamiltonian studied by Friedrich and
  Wintgen~\cite{WF,FW},
 \begin{equation}
 H(\mathbf x,\mathbf p) = {\mathbf p^2 \over 2m} - {e^2\over r} +
 {\textstyle\frac12} m\varpi^2 (x^2+ y^2).
  \label{7.6}\end{equation}
 It describes, in a rotating coordinate system, a hydrogen atom in 
a constant magnetic field
 $B = (2mc/e)\varpi$ along the $z$~axis.
 This is
 ``a real physical system that can be and has been studied in the 
laboratory''\negthinspace.
 The magnetic field is a continuous variable that is under the 
experimenter's control;
 the charge of the proton is~not!
 Therefore, 
to get a family of experimentally realizable systems
one applies (\ref{7.1}), (\ref{7.4}), (\ref{7.5})
  with $\nu_1= -1$, $\nu_2= 2$.
 Thus 
 \begin{equation}
 \lambda_2 = \alpha^{-6} \lambda_{20}\,, \qquad
 E = \alpha^{-2} E_0\,, 
 \label{7.7}\end{equation}
 and the bottom-line equation (\ref{6.6}) becomes
 \begin{equation}
 \rho_\gamma(E) = a_\gamma \sin\left[{S_0\over \hbar} 
 \left( \frac E{E_0}\right)  ^{-1/2} + \eta_\gamma \right].
  \label{7.8}\end{equation}
 In the  notation of Friedrich and Wintgen,
 \begin{equation}
 \Gamma \equiv {\varpi \over \varpi_0} = \alpha^{-3}\,, \qquad
\frac E{E_0} = \Gamma^{2/3},
  \label{7.9}\end{equation}
 and the contribution $\rho_\gamma$ to the spectral density
 is a sinusoidal function of $\Gamma^{-1/3}$.
 (We  write  $\Gamma$ and $\varpi$ for the $\gamma$ and $\omega$ 
 of \cite{FW} to avoid notational collisions.)
For the beautiful fruits of this approach as applied to (\ref{7.6})
  we refer to \cite{WF,FW,HRW}.

 From $S= \alpha^{\nu_1+2}S_0$ and $E= \alpha^{2\nu_1} E_0$
  we still get
 \begin{equation}
 \frac{d S}{dE} = {\nu_1+ 2 \over 2\nu_1}\, \frac SE\,,
  \label{7.10}\end{equation}
 but this is not equal to an orbital period in these incompletely 
homogeneous systems.
 
More generally,
 let us repeat the analysis of (\ref{7.1})--(\ref{7.5}), but with 
an arbitrary potential in the role of~$V_2\,$.
 Instead of a scaling of $\lambda_2$ as in (\ref{7.5}), one 
obtains a one-parameter family of potentials 
 \begin{equation}
 V_{2[\alpha]}(\mathbf x) \equiv \alpha^{2\nu_1} V_2(\alpha^{-2} 
\mathbf x).
 \label{7.11} \end{equation}
 The classical orbits persist as $E$ and $\mathbf x$ rescale
  and $V_2$ deforms in this way, while the homogeneous potential 
  $V_1$ and its coupling constant remain unchanged. 
When $V_2$ is a sum of several homogeneous terms with different 
 exponents~$\nu_j$,
 a separate scaling law for each coupling constant,
 \begin{equation}
\lambda_j = \alpha^{2(\nu_1-\nu_j)} \lambda_{j0}\,,
\label{7.12} \end{equation}
results from (\ref{7.11}).
This last situation has been discussed thoroughly by 
 Friedrich \cite{Fr1,Fr2}, with special attention to the 
easily  experimentally realizable case where $V_1$ is a Coulomb 
potential and the $\lambda_j$ are the strengths of applied constant 
magnetic and electric fields.

 \section{Implications and summary} \label{sec:imp}

 A central feature of the relation between classical and quantum 
mechanics is that quantization introduces into each problem a new 
fundamental scale, set by the quantum of action,~$\hbar$.
 That is, one-parameter families of situations that are equivalent 
classically become distinct quantum-mechanically.
  Taking the semiclassical limit refers to motion along one of these 
families in a certain direction, the opposite direction leading to 
``deep quantum'' behavior.
 Moving from one family to another, on the other hand, corresponds 
to various purely classical distinctions, such as integrable versus 
chaotic, or different values of angular momentum.

 When the potential is homogeneous (Sec.~\ref{sec:hom})
  or absent (Sec.~\ref{sec:bill}), a 
``situation'' simply means a point 
 in phase space.
  In the more general contexts of Secs.\ \ref{sec:coup} 
  and~\ref{sec:mix}
 the space of situations should be enlarged by one dimension, 
representing the coupling constant $\lambda$ or $\lambda_2$
or the parameter $\alpha$ in~(\ref{7.11}).
In either case the points are grouped into equivalence classes, 
the classical trajectories.
Finally, the
trajectories (including, in particular, the closed orbits) 
fall into families, (\ref{5.4}) or (\ref{6.2}),
  related by a geometrical 
similarity and therefore mathematically equivalent.
 The classical trajectories of a family  are exactly the same
 at all energies, except for a trivial rescaling. But the 
quantum states at various energies are quite different.

  In the study of spectra the most convenient choice of parameter
 along each family is  the energy.
 What determines whether an energy is large or small?
 To compare it with $\hbar$ one must construct a quantity with 
dimensions of action.
 The dimensions of $\sqrt{2m|E|}$ are action divided by length,
 so a suitable measure is the product of $\sqrt{2m|E|}$
  with some length characteristic of the entire system.
The statement ``$\hbar$ is small'' is meaningless unless translated 
into such a criterion.
 In the end one can always choose units in which $\hbar=1$
 and $2m=1$, 
 and then the only independent physical dimension is length.
 (Once the quantum of action and the characteristic length of the 
system are fixed, rescaling $m$ amounts to changing the unit of 
time.
 Since $m$ has been totally inert throughout  our 
considerations, it could have been eliminated 
at the very beginning, but so that the classical-mechanical 
equations would look familiar, we did not do so.)
 The criterion now is  whether the length 
 $|E|^{-1/2}\equiv \omega^{-1}$ 
 is large or small relative to the length scale of the system.
 This restatement makes sense in classical wave theories, such as 
optics or acoustics, as originally studied by Balian and
  Bloch~\cite{BB1,BB3}.

The length scale can be set by the geometry of the boundary, 
 if there is one.
 Otherwise it must be a characteristic of the potential.
 At first sight this is a simple matter:
Quantities such as $\nabla^2V/V$  (evaluated at a minimum of the 
potential, say)
 are independent of the coupling 
constant and characterize the spatial scale of the potential, as 
distinct from its strength.
 However, if $V$ is homogeneous, (\ref{6.1}) shows that the 
distinction between the spatial scale and the coupling constant is 
a mirage; 
 and Sec.~\ref{sec:mix} suggests that this ambiguity can be 
imported into more general potentials as well. 
 For a homogeneous potential there is an alternative way to set the 
scale of the system, exploiting (\ref{6.1}) to ``transmute'' the 
coupling constant into a length, $x_0 \equiv \beta^{-1}$. 
 For example, if $d=1$ and the potential energy is $\lambda x^\nu$,
 then $\lambda$ has dimensions $\text{[energy][length]}^{-\nu}$
 and can be written
 \begin{equation}
 \lambda \equiv {\hbar^2\over 2m} \, x_0{}\!^{-(\nu+2)},
  \label{8.1}\end{equation}
 so that
 \begin{equation}
 \lambda x^\nu = {\hbar^2\over 2m x_0{}\!^2}
 \left ({x\over x_0}\right) ^\nu. 
\label{8.2}\end{equation}

\begin{figure}
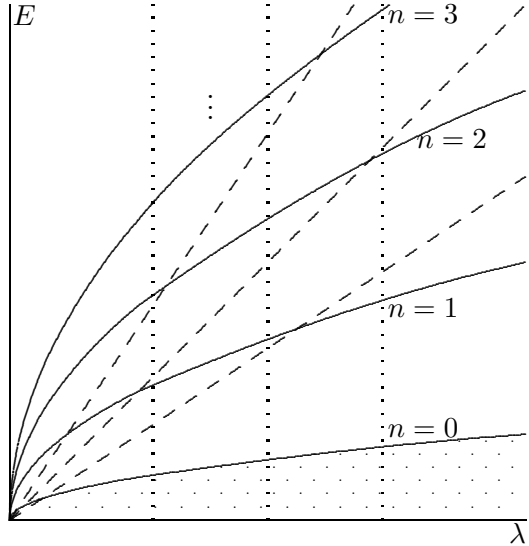

 $$\beginpicture 
\setcoordinatesystem units <.3truein, .3truein>
 \setplotarea x from 0 to 9, y from 0 to 9
 \putrule from 0 0 to 0 9
 \putrule from 0 0 to 9 0
 \put{$E$} [lt] <1pt,0pt> at 0 9
 \put{$\lambda$} [rt] <0pt,-1pt> at 9 0
 \put{$\vdots$} [b] at 3.5 7 
 \put{$n=0$} [lb] <2pt,1pt> at 6.5 1.4
  \put{$n=1$} [lb] <2pt,1pt> at 6.5 3.5
 \put{$n=2$} [lb] <2pt,0pt> at 7 6.5
 \put{$n=3$} [lt] <2pt,0pt> at 6.5 9
  \setquadratic
  \vshade
 0 0 0 
 .25 0 .25
 1 0 .5
  4 0 1
  9 0 1.5 /
 \plot
 0 0
 .0625 .125
 .25  .25
 .49 .35
 .74 .43
 1.44   .6
 2.25  .75
 5.76 1.2
 9 1.5 /
 \plot 
 0 0
 .0625 .375
 .25 .75
 .49 1.05
 .74 1.29
 1.44 1.8
 2.25 2.25 
 5.76 3.6
 9 4.5 /
 \plot
 0 0
 .0625 .625
 .25 1.25
 .49 1.75
 .74  2.15
 1.44  3
 2.25 3.75
 5.76 6
 9 7.5 /
 \plot
 0 0
 .0625 .875 
 .25 1.75
 .49 2.45
 .74 3.01
 1.44   4.2
 2.25   5.25 
 4 7
 6.63  9 /
 \setlinear 
 \setdashes
 \plot
  0 0 
 9 9 /
 \plot
 0 0 
 9 6 /
 \plot
 0 0
 6 9 /
 \linethickness=1pt
 \setdots
 \putrule  from 2.5 0 to 2.5 9
 \putrule from 4.5 0 to 4.5 9
 \putrule from 6.5 0 to 6.5 9
\endpicture$$
\caption{Solid curves:
  Energy levels of the harmonic oscillator,
 $E_n = \bigl(n+\frac12\bigr)\hbar\varpi$,
  as functions of coupling 
constant, $\lambda\propto \varpi^2$.
No quantum states exist in the shaded area.
 Dashed lines: Loci of typical classical orbits fixed in space 
(Sec.~\ref{sec:coup}). 
 Dotted lines: Loci of typical homogeneously scaled orbits
 (Sec.~\ref{sec:hom}).
 The direction of increasing $E$ along either set of lines leads 
into the semiclassical region of large quantum numbers.}
\label{scalharm}
\end{figure}

\begin{figure}
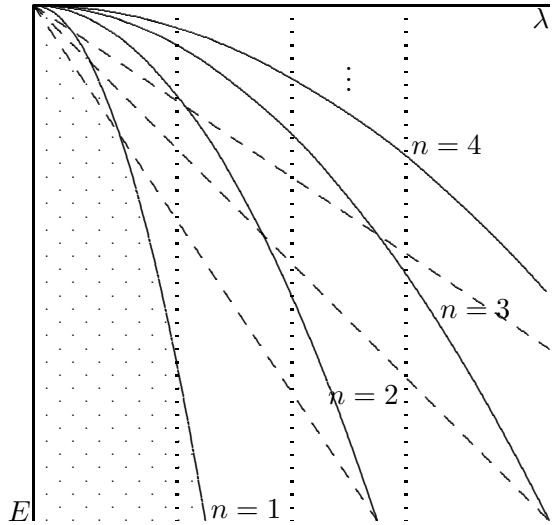

$$\beginpicture 
\setcoordinatesystem units <.3truein, .3truein>
 \setplotarea x from 0 to 9, y from -9 to 0
 \putrule from 0 0 to 0 -9
 \putrule from 0 0 to 9 0
  \put{$E$} [rb] <-1pt,0pt> at 0 -9
 \put{$\lambda$} [rt] <0pt,-1pt> at 9 0
 \put{$\vdots$} [b] at 5.5 -1.5 
  \put{$n=4$} [lb] <2pt,0pt> at 6.5 -2.6
  \put{$n=3$} [lb] <2pt,1pt> at 7 -5.5
 \put{$n=2$} [lb] <3pt,1pt> at 5 -7
 \put{$n=1$} [lb] <2pt,1pt> at 3 -9
 \setquadratic
  \hshade
 -9  0 3
-4  0  2 
-1 0 1
-.64 0 .8
-.25 0 .5
  -.0625 0 .25
 0 0 0 /
 \plot
 0 0
.25   -.0625
 .5 -.25
.7   -.49
 .8 -.64
 .9 -.81
1  -1 
1.414 -2
 1.732 -3
 2 -4
 2.236 -5
 2.449 -6
 2.646 -7
 2.828 -8
 3 -9  
  /
 \plot 
 0 0
 .5 -.0625
1  -.25
 1.4 -.49
1.6  -.64
 1.8 -.81
2  -1 
2.828 -2
3.464 -3
 4 -4
 4.472 -5
 4.899 -6
 5.292 -7
 5.657 -8
 6 -9 
 /
  \plot
 0 0
 .75 -.0625
1.5   -.25
 2.1 -.49
2.4 -.64
 2.7 -.81
3  -1 
4.243 -2
5.196 -3
 6 -4
 6.708 -5
 7.348 -6
 7.937 -7
 8.485 -8
 9 -9 
 /
  \plot
 0 0
 1 -.0625
 2 -.25
2.8   -.49
 3.2 -.64
 3.6 -.81
4  -1 
5.657 -2
6.928 -3
 8 -4
 8.944 -5
  /
 \setlinear 
 \setdashes
 \plot
  0 0 
 9 -9 /
 \plot
 0 0 
 9 -6 /
 \plot
 0 0
 6 -9 /
 \linethickness=1pt
 \setdots
 \putrule  from 2.5 0 to 2.5 -9
 \putrule from 4.5 0 to 4.5 -9
 \putrule from 6.5 0 to 6.5 -9
\endpicture$$
\caption{Solid curves:
  Energy levels of the hydrogen atom,
 $E_n \propto - e^4/n^2$, as functions of coupling 
constant, $\lambda\propto e^2$.
No quantum states exist in the shaded area.
 Dashed lines: Loci of typical classical orbits fixed in space 
(Sec.~\ref{sec:coup}). 
 Dotted lines: Loci of typical homogeneously scaled orbits
 (Sec.~\ref{sec:hom}).
  The semiclassical region of large quantum numbers
 is reached by moving 
 along the dashed lines in the direction of
 increasing $|E|$ ($E\to -\infty$), 
 or
 along the dotted lines in the direction of
 decreasing $|E|$ ($E\to 0$).}
\label{scalcoul}
\end{figure}

  It is instructive to look at everyone's favorite homogeneous 
  potentials, the harmonic oscillator (\ref{4.11})
  and the hydrogen atom (\ref{6.11}),
  in the light of the foregoing remarks.
 If we treat these systems according to Sec.~\ref{sec:coup},
 we are taking a fixed ``classical situation'' to be a fixed ratio
 $E/\lambda$, with $\lambda\propto \varpi^2$ or $e^2$ respectively.
 We expect to encounter semiclassical behavior as $|E| \to \infty$ 
along one of the diagonal lines in Figs.\ 
 \ref{scalharm} and~\ref{scalcoul},
 and indeed this leads into the region of large quantum numbers
 in either case.
There is no unique way to associate a natural length with either of 
these potentials, but lengths independent of~$\lambda$ can be 
 built out of derivatives of~$V$ at an arbitrarily chosen point.
 If we treat the systems according to Sec.~\ref{sec:hom}, 
  we keep $\lambda$ constant and therefore move along one of the 
  vertical lines in the figures.
 From
 \begin{equation}
 x_0 = \lambda^{-1/(\nu+2)}  \qquad (\hbar^2= 2m),
 \label{8.3}\end{equation}
  we have
 \begin{equation}
 x_0{}\!^2 = \frac1\varpi\,,
  \qquad E_nx_0{}\!^2 = n+\frac12
  \label{8.4}\end{equation} 
 for the energy levels of the oscillator, and 
 \begin{equation}
 x_0{}\!^2 =e^{-4}, \qquad 
 E_nx_0{}\!^2 \propto - \,\frac1{n^2}
 \label{8.5}\end{equation} 
 for those of the atom.
 The semiclassical regime of large quantum number~$n$ is reached by 
going vertically upward in either of the figures.
For the oscillator this is again the limit of large~$E$, albeit in 
a different direction in the $E$--$\lambda$ plane.
 But for the Coulomb potential this limit corresponds to $E \to 0$,
 not $|E|\to \infty$.
 (In view of (\ref{6.13}) this is still the direction of 
increasing~$S$.) 
 This (known) result is  not  surprising in view of the 
negative exponent in (\ref{6.6}), (\ref{6.13});
 but it indicates that the identification of ``semiclassical'' with 
``large energy'' can be ambiguous, even within the theory of the 
same physical system.

\goodbreak
 Let us summarize the main points:
 \begin{enumerate}
 \item  
 The concept of ``spectral oscillations'' with respect to
energy (associated with the basic formula (\ref{3.2}), which we
took as given) is problematical, because the classical orbit
structure itself varies with the energy, quantitatively and
sometimes qualitatively.

 \item {\em The orbits can be ``nailed down'' by scaling the 
coupling constant along with~$E$.}
 From this enlarged perspective a mechanical system (with 
varying~$\lambda$) has one (very clearly distinguished)
  ``quantum dimension'' along which very well-defined
 spectral oscillations take place.
It is necessary to study an entire family of systems 
(parametrized by~$\lambda$) to gain a clear and complete picture of a
single system; this is usually not a problem theoretically, though it may
be very hard experimentally.

 \item For a homogeneous or partially homogeneous potential
there are other options for fixing the orbit structure, involving
dilations.

 \item The wave frequency $\omega = E^{1/2}$ is a better variable 
than $E$ for describing the oscillations globally. 
 (In the homogeneous case the exponent $\frac12$ is replaced by 
 $(\nu+1)/ 2\nu$.)
 Correspondingly, the best descriptor of the oscillations 
caused by a particular orbit 
 is a characteristic length (\ref{5.10}) of the orbit, 
 not the orbital period.

 \item Even the local frequency of spectral oscillation with 
respect to energy,
 $dS/dE$, is not equal to the orbital period when the coupling 
constant is varied.
 (In passing we supplied a proof of the equality when $\lambda$ is 
constant.)
\end{enumerate}

    Points 1 and 2 require further comment in view of the fact
that, strictly speaking, the Gutzwiller formula (\ref{3.2})
applies only to an orbit that is both isolated and unstable.
Similar formulas apply when the orbits form continuous families
because of symmetries \cite{CL1,CL2}, with completely integrable
systems \cite{BT2} as the extreme case.  Those formulas have the
same simple oscillatory nature as (\ref{3.2}), but with different
phases and amplitudes; our observations apply to them unchanged.
(Indeed, the diamagnetic Kepler problem (\ref{7.6}) is of this
type, when it has not been reduced to an angular momentum
eigenspace and a two-dimensional classical configuration space.)
Complications develop, however, when symmetric systems are
slightly perturbed \cite{Cr,UGT,Sie2} or when {\em stable\/} orbits
develop  bifurcations \cite{Ri,OH,Sie1}.  In
these cases the semiclassical formulas involve factors (typically
Bessel functions) that interpolate between different oscillatory
regimes.  These phenomena occur even in billiard systems, as
pointed out in \cite[Appendix~D]{Sie1}  and exemplified in
detail in various studies of the ellipse \cite{Cr,Sie2,WWD}.
Therefore, the observations of the present work provide no escape
from them.  The point is, however, that {\em whatever\/} takes
the place of (\ref{3.2}) must scale in the same simple way when
the energy and coupling constant are simultaneously scaled as in
Sec.~V{}. This is so because the semiclassical formulas are built
entirely out of classical ingredients, along with strategically
placed factors of $\hbar$ to give all quantities the correct
dimensions.
 
    Two  examples are Sieber's formula for the ellipse
\cite[eq.~(64)]{Sie2} and the same author's generic formula for
bifurcations \cite[eq.~(37)]{Sie1}.  In the former case, given a
fixed ellipse the only relevant parameter is
$\omega\equiv\sqrt{E}$ ($k$, in the notation of \cite{Sie2}).
The formula describes a rigid function of $\omega$ just like our
(\ref{5.8}), albeit of a more complicated functional form.  In
the other case, in contrast, two things are going on
simultaneously.  The energy is changing, but also the function
being evaluated upon the energy is changing.  (Most
spectacularly, some terms completely disappear when the energy
passes through the bifurcation point, where two ``satellite
orbits'' disappear from the classical structure.)  A bifurcation
also occurs in the billiard problem, but one must vary the
eccentricity of the ellipse to see it.  {\em By scaling the
coupling constant along with the energy, one introduces a
``normal coordinate system'' into the manifold of problems under
study, so that the quantum parameter $\omega$ is decoupled from
the classical parameters defining the system.}  In the ellipse
problem this was automatic:  the original energy is already a
good quantum parameter, and the eccentricity is the most
important classical parameter.  For a Schr\"odinger problem
(\ref{5.1}) the normal coordinates are less natural:  the quantum
parameter is the joint magnitude of $E$ and $\lambda$; varying it
leaves the system at a fixed effective ``distance'' from the
bifurcation; and letting $\lambda$ deviate from the scale of $E$
amounts to changing the classical mechanics toward or away from
the bifurcation.  Although this approach will not by itself play
the major role in deriving more formulas like \cite[(37)]{Sie1},
we submit that it can aid considerably in their interpretation.

 \section{A final comment} \label{sec:reso}

 When periodic-orbit theory was first applied numerically to the 
spectra of concrete systems, early authors expressed surprise that 
it gave accurate results ``even'' for the lowest eigenvalues.
 As a semiclassical method, the technique had been expected to be 
applicable primarily in the regime of ``large quantum numbers'' and 
hence high energy.
 Later it became clear that the method was generally practical 
 {\em only\/} in the low-energy regime, because the number of 
periodic orbits increases roughly exponentially with period,
 and as energy increases it becomes necessary to consider 
increasingly long orbits in order to resolve individual  
eigenvalues~\cite{Berry,Bolte}.
 Long after this fact has been accepted, it is still often regarded 
as ``paradoxical''\negthinspace.

 We suggest that, in hindsight, this phenomenon is an 
instance of the familiar principle that when a calculation is very 
stable and involves some kind of smoothing or averaging, the 
inverse calculation is likely to be unstable (highly dependent on 
the details of the input) and hence difficult. 
(The following remarks obviously lack technical precision and are
offered as preliminary ideas only.  They do not explain how
formulas such as (\ref{3.2}) apparently have a wider validity
than the stationary-phase arguments that lead to them.)
 There is a loose analogy with the solution of an initial-value 
problem for the heat equation.
For large positive time the problem is very easy to solve and very 
insensitive to the details of the initial data; 
 consequently, reconstructing the data from the final solution is 
hopeless.
 For very small time an approximate solution to the backwards 
problem, adequate for some practical purpose,  may be feasible; 
of course, the forward problem is less trivial in that case.

In spectral asymptotics we are interested in deducing the spectrum 
of a differential operator from its geometry,
 or vice versa.
 Here ``geometry'' is meant in an extended sense, including not 
only the literal geometry of the region $\Omega$ where the wave 
functions are defined, but also the potential function
 (or other coefficient functions in the operator~$H$).
 The classical periodic orbits of the Hamiltonian
  $H(\mathbf x,\mathbf p)$ 
are also aspects of the geometry of the system.

 In the ``old'' spectral asymptotics, associated with the names 
Weyl, Thomas--Fermi, Minakshisundaram, Schwinger--DeWitt, etc.,
one associates the (high-$E$) asymptotic behavior of the density of 
states with the global geometry of the operator: the volume of 
$\Omega$, the integrated curvature of its boundary, the integral of 
the Ricci curvature of $\Omega$ if it is a manifold,
  the integral of the potential $V$ over $\Omega$, and so on. 
 Via the asymptotics of the heat kernel, the passage from the 
spectrum to the geometry is rigorously asymptotic.
 In the inverse direction, however, the geometry does not determine 
a genuine asymptotic expansion of the eigenvalue density in powers 
of~$E^{-1}$, precisely because of the presence of the oscillatory 
terms that are the subject of the periodic-orbit theory. 
 The formal expansion of the eigenvalue density becomes literally 
asymptotic only when some kind of averaging is performed, such as 
Lorentzian smoothing~\cite{BB1} or Riesz means~\cite{Hor}.

 A complementary situation has always existed at the low end of the 
spectrum.
 If we know $\Omega$ and $H$ exactly, then it is relatively easy to 
construct the lowest-lying eigenvalues and eigenfunctions, by
 variational methods, numerical methods, etc.
 Larger eigenvalues are harder.
 Furthermore, merely from a knowledge of the lowest energies one 
would not expect to be able  to reconstruct $\Omega$ or~$H$.
 (Complete knowledge of a single eigen{\em function\/} is a 
different matter, however.) 

 The ``new'' spectral asymptotics of periodic orbits extends this 
picture.
Decreasing the width over which the 
eigenvalue distribution is averaged,
  one supplements the power-law asymptotics 
with the longest-wavelength oscillatory components (\ref{3.2}), 
characterized by the lengths or periods of the shortest classical 
orbits.  This is still an asymptotic (high-energy) matter.
 It is best thought of as a prediction of classical behavior
 (well-defined orbits for wave packets) in the high-energy, 
 large-action regime.
 It still involves an averaging of the spectrum, albeit on a 
smaller scale.
 The stable direction of prediction is {\em from\/} the spectrum 
 {\em to\/}  the classical, geometrical description.

The counterpart of this at low energy is that from the classical 
orbits 
 (and their associated amplitudes and phases, cf.~(\ref{3.2}))
one can predict the formation of discrete eigenvalues
 (resonant frequencies), with 
greatest precision near the bottom of the spectrum.
 In some  sense this involves an averaging over geometrical 
information.  
 More precise geometrical information in principle allows precise 
construction of longer-period orbits and hence more complete
reconstruction of the spectrum (both improved resolution 
 and extension to higher energies).
 The stable direction of prediction is {\em from\/} the 
 geometry {\em to\/}  the spectrum.
(In the other direction, ``[M]issing a level has disastrous 
consequences for locations, heights, and even the very appearance 
of the non-Newtonian peaks in the Fourier transform.'' 
\cite{BBSKB})

 It is misleading to think of the regime of low energy 
 (or small quantum numbers) 
 as the ``deep quantum regime''\negthinspace, 
 as if the duality between the spectrum of quantum 
eigenvalues and the spectrum of classical 
periodic orbits is irrelevant there.
A better phrase is ``resonant regime'':
There, resonant behavior emerges on the spectral side, out of the 
geometry and classical mechanics, 
  much as, in the opposite ``classical regime''\negthinspace,
classical-mechanical behavior emerges on the geometrical side, out 
of the quantum substrate.

 \begin{acknowledgments}
 The calculation in Sec.~\ref{sec:peri}
  developed in conversations with David Garfinkle
  at the Parkerfest (Oakland University, 2000).
 I thank Y. Colin de Verdi\`ere,   F. H. Molzahn, Y. Dabaghian, and
 the referees for helpful comments and references.

\end{acknowledgments}

\goodbreak

 \end{document}